# THE PRINCIPAL PARAMETERS OF UNSTUDIED OPEN CLUSTERS WITH *NIR* OBSERVATIONS

Tadross, A. L. & Nasser, M. A.

*National Research Institute of Astronomy & Geophysics, NRIAG, Cairo, Egypt*

**ABSTRACT**

We studied the principal parameters of some previously unstudied open star clusters using the *JHK* Near-IR photometry (*2MASS survey*). These clusters have been selected from the updated Catalogs of Dais and Webda. Based on the *2MASS* database and the *DSS* visual images, some homogeneous methods and algorithms have been applied. The astrometry and photometric principal parameters are determined for the first time.

## 1. INTRODUCTION

Open clusters are very important tools in studying the formation and evolution of the Galactic disk, the star clusters themselves, and the stellar evolution. The fundamental physical parameters of open clusters, e.g. distance, reddening, age and metallicity, are very necessary for studying the Galactic disk. The Galactic (radial and vertical) abundance gradient also can be studied by open clusters (Hou et al. 2000; Chen et al. 2003; Kim & Sung 2003; Tadross 2003; Kim et al. 2005). They are excellent probes of the Galactic disc structure (Janes & Phelps 1994; Bonatto et al. 2006). The strong interest of open clusters results come from their fundamental properties. The stars of a cluster, which are lie at the same distance from the Sun, have the same chemical composition and the same age. According to some estimation, there are as many as 100,000 open clusters in our Galaxy, but less than 2000 of them have been confirmed and cataloged.

Dolidze clusters No. 9, 10, 11, 19, 21, 26, 27, 37, 39, and 41 are not studied before. There are no information known about those 10 clusters except their coordinates and apparent visual diameters, see Table 1. The photometric properties for them have been studied here for the first time using 2MASS database and DSS catalog.

The most important thing for using *NIR* database (*2MASS*) is the powerful detection of the stars behind the hydrogen thick clouds, which act as curtains, on the galactic plane, see the example in Fig 1.

Table 1: The coordinates and apparent diameters of the clusters under investigation.

| Cluster | RA $^{h\ m\ s}$ | Dec $^{\circ\ '\ ''}$ | G. long. $^\circ$ | G. lat. $^\circ$ | Diameter ' |
|---|---|---|---|---|---|
| Dolidze 9 | 20 25 42 | +41 56 00 | 79.898 | 2.293 | 5 |
| Dolidze 10 | 20 26 18 | +40 07 00 | 78.481 | 1.150 | 3 |
| Dolidze 11 | 20 26 30 | +41 27 00 | 79.590 | 1.892 | 5 |
| Dolidze 19 | 05 23 42 | +08 11 00 | 195.104 | -15.317 | 23 |
| Dolidze 21 | 05 27 24 | +07 04 00 | 196.585 | -15.103 | 12 |
| Dolidze 26 | 07 30 06 | +11 54 00 | 206.508 | 13.901 | 23 |
| Dolidze 27 | 16 36 30 | -08 57 00 | 7.616 | 24.664 | 25 |
| Dolidze 37 | 20 03 00 | +37 41 00 | 73.941 | 3.576 | 8 |
| Dolidze 39 | 20 16 24 | +37 52 00 | 75.536 | 1.453 | 12 |
| Dolidze 41 | 20 18 49 | +37 45 00 | 75.707 | 0.992 | 11 |

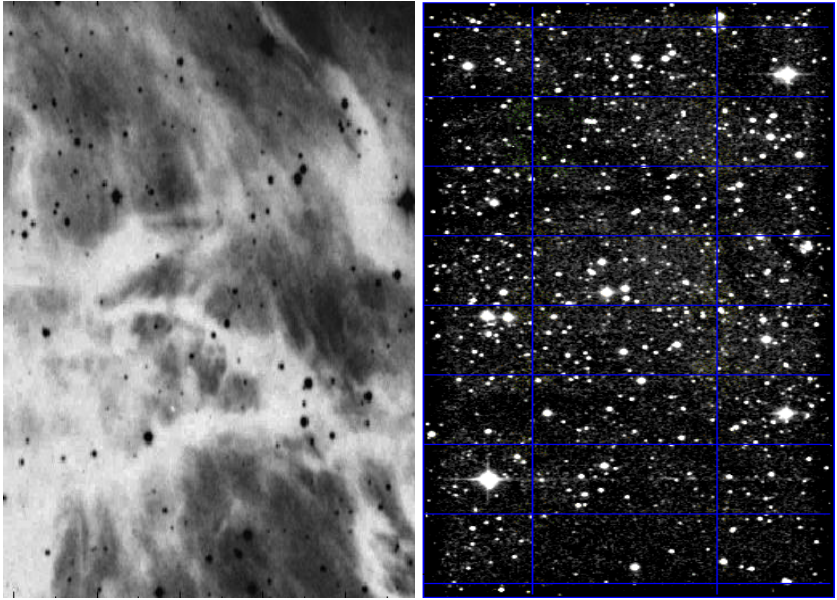

Fig 1: A comparison between the *DSS* visual (left panel) and *2MASS* (right panel) images of open star cluster "Dolidze 10".

## 2. DATA REDUCTIONS

The photometric data of the investigated clusters have been obtained from the Two Micron All Sky Survey (2MASS), Skrutskie et al. (1997), which are available at: ***http://www.ipac.caltech.edu/2mass/releases/allsky/***

2MASS provides J, H, and $K_s$-band photometry for millions of galaxies and nearly a half-billion stars (Carpenter, 2001).

2MASS observations are obtained using two highly 1.3-m telescopes, one at Mount Hopkins in Arizona (for northern survey) and the other at Cerro Tololo in Chile (for southern survey). Each telescope was equipped with three-channel camera, each channel consisting of 256 x 256 array HgCdTe detectors. It is uniformly scanning the entire sky in three near-IR bands J (1.25 μm ), H (1.65 μm) and $K_s$ (2.17 μm).

2MASS has proven to be a powerful tool in the analysis of the structure and stellar content of open star clusters (Bica et al. 2003, Bonatto & Bica 2003). From Soares & Bica (2002), we can see that the errors are more affect for $K_s$ at given magnitude. So J and H data have been used here to probe the faint stars of these clusters with more accuracy.

Data extraction has been performed using the known tool of VizieR for 2MASS data base. The clusters' data extracted at a preliminary radius of 30 arcmin from their obtained centers. The nominated clusters should have enough members, prominent main sequences, and good images on the Digitized Sky Survey (DSS). Cut-off of photometric completeness limit at J < 16.5 mag is applied on the 2MASS data to avoid the over-sampling at the lower parts of their CMDs (cf. Bonatto et al. 2004 & Tadross 2008). Stars with observational uncertainties of J, H, K > 0.20 mag have been eliminated. Membership criteria are adopted for the location of the stars in the CMDs' curves within 0.10 - 0.15 mag around the ZAMS.

### 3. Clusters' Diameters Determination

To determine the radius a cluster, the radial surface density of that cluster must be achieved firstly. The cluster border is defined as the surface which covers the entire cluster area and reach enough stability in the background density, i.e., the difference between the observed density profile and the background one is almost equal zero (cf. Tadross 2004). The determination of a cluster radius is made possible by the spatial coverage and uniformity of 2MASS photometry which allows one to obtain reliable data on the projected distribution of stars for large extensions around the clusters' center, Bonatto et al. (2005). The fitting of King (1962) has been applied to derive the cluster's radius, as shown for the example in Fig 2.

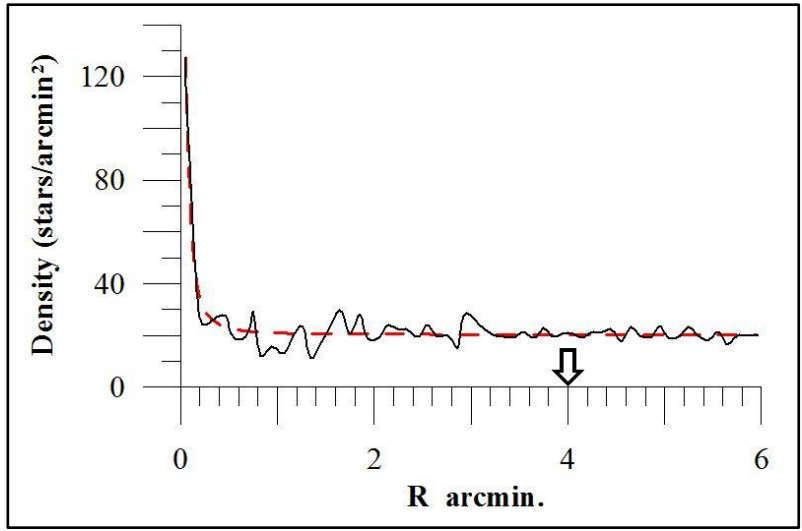

**Fig 2: An example for the lower radius determination for "Dolidze 39", the curved dotted line represents the fitting of King (1962). The arrow refers to that point, at which the radius obtained; reaching the background density of about 20 stars per arcmin$^2$.**

## 4. PHOTOMETRIC ANALYSIS

The main parameters for these clusters (age, reddening, and distance) can be determined by fitting the solar type isochrones of *Padova* (Bonatto et al. 2004) to the CMD of each cluster. Several fittings have been applied to each cluster using different age's isochrones. CMDs have been constructed for our candidates; the stars located away from the main sequences are excluded (cf. Bonatto et al. 2005, Tadross 2008 and references therein). Once the best fit has been obtained, the age, distance, and color excess $E_{J-H}$ can be estimated. The color excess $E_{B-V}$ can be also determined; applying Dutra et al. (2002)'s relations.

Once the cluster's distance is estimated, distance from the galactic center, $R_{gc}$, and the projected geometric distances on the galactic plane from the Sun, $X_\odot$, $Y_\odot$, $Z_\odot$ (the distance from galactic plane), and also the linear diameters of the clusters can be determined. Table 2 presents all the clusters' properties that estimated in the present work. The *2MASS* images and CMDs of each cluster can be seen in Figs 3.

Table 2: The main parameters of the clusters; Columns from left to right represent the cluster's name, angular diameters, linear diameters, age, reddening ($E_{J-H}$ & $E_{B-V}$), distance modulus, distances from the sun in parsecs; the projected distances on the galactic plane from the Sun, $X_\odot$, $Y_\odot$, $Z_\odot$ in parsecs; and the distance from the galactic center in kilo parsecs, respectively.

| Name | Diam. arcmin | L. Diam. pc | Age Myr | E(J-H) mag | E(V-B) mag | m-M mag | Dist. pc | $X_\odot$ pc | $Y_\odot$ pc | $Z_\odot$ pc | $R_{gc}$ kpc |
|---|---|---|---|---|---|---|---|---|---|---|---|
| Dolidze 09 | 6.0 | 1.5 | 20 | 0.25 | 0.80 | 10.4 | 866 | -152 | 852 | 35 | 8.4 |
| Dolidze 10 | 3.6 | 1.0 | 250 | 0.25 | 0.80 | 10.6 | 950 | -190 | 931 | 19 | 8.4 |
| Dolidze 11 | 5.0 | 1.6 | 400 | 0.26 | 0.83 | 11.0 | 1127.0 | -204 | 1108 | 37 | 8.4 |
| Dolidze 19 | 24.0 | 9.2 | 160 | 0.35 | 0.55 | 11.6 | 1320 | 1229 | -332 | -349 | 9.8 |
| Dolidze 21 | 12.0 | 5.1 | 200 | 0.35 | 0.55 | 11.8 | 1447 | 1339 | -399 | -377 | 9.9 |
| Dolidze 26 | 22.0 | 12.2 | 100 | 0.28 | 0.44 | 12.2 | 1907 | 1657 | -826 | 458 | 10.2 |
| Dolidze 27 | 28.0 | 8.3 | 50 | 0.48 | 0.75 | 11.4 | 1015 | -914 | 122 | 424 | 7.5 |
| Dolidze 37 | 8.0 | 3.5 | 300 | 0.31 | 0.48 | 11.7 | 1490 | -411 | 1429 | 93 | 8.2 |
| Dolidze 39 | 8.0 | 2.0 | 250 | 0.28 | 0.44 | 10.5 | 872 | -218 | 844 | 22 | 8.3 |
| Dolidze 41 | 10.0 | 5.1 | 400 | 0.34 | 0.53 | 12.2 | 1763 | -435 | 1708 | 31 | 8.2 |

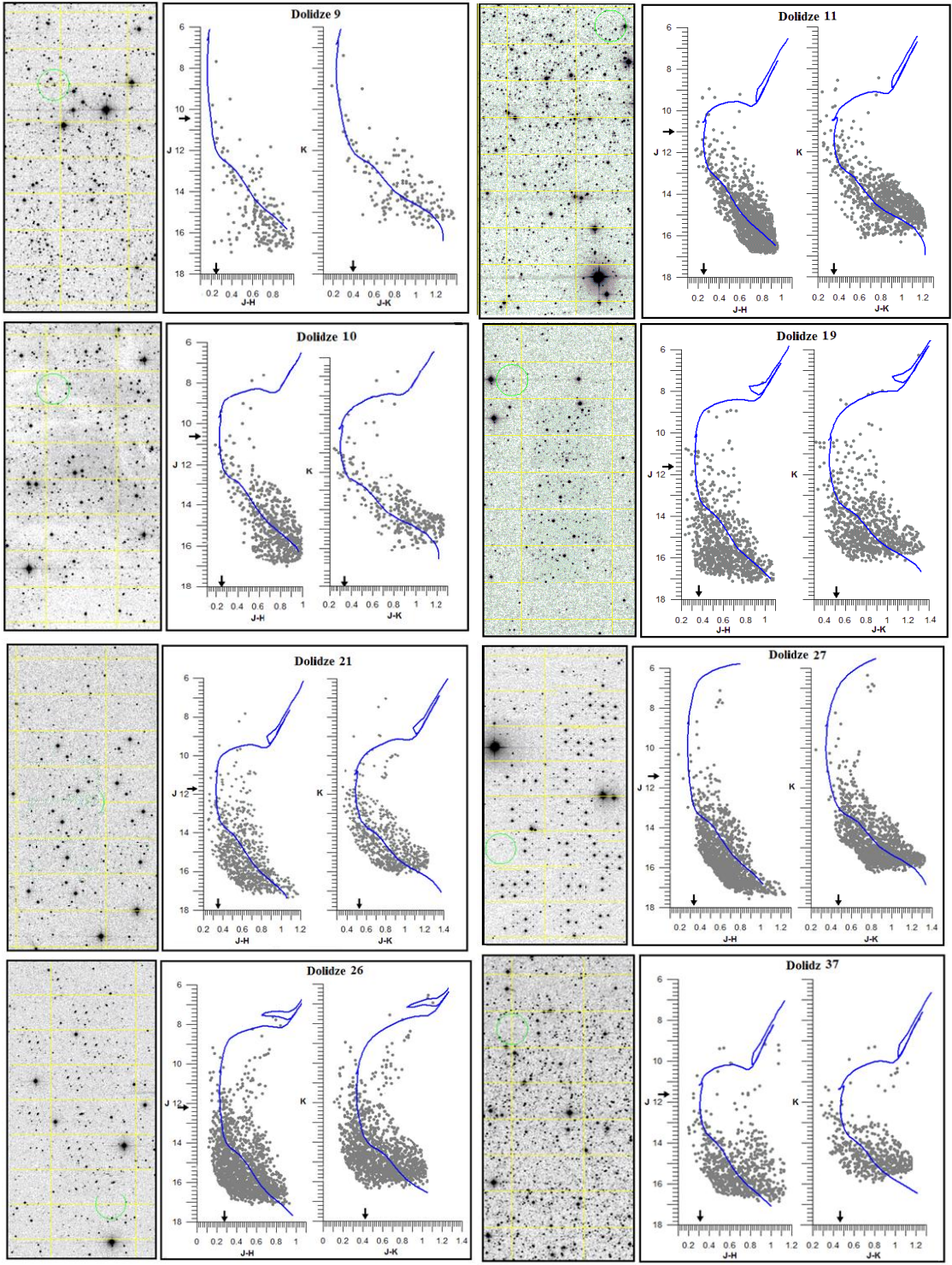

**Fig 3:** *2MASS* images and CMDs of Dolidze 9, 10, 11, 19, 21, 26, 27, and Dolidze 37 respectively.

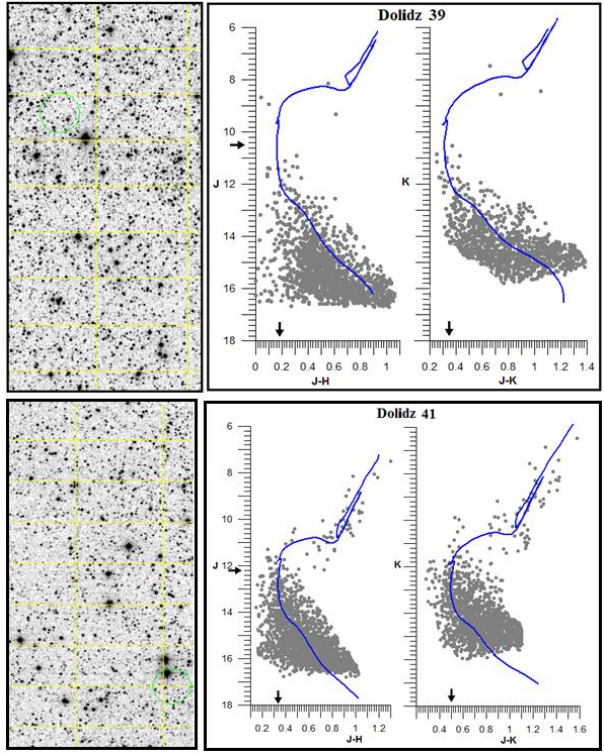

**Fig 3- Continued:** *2MASS* **images and CMDs of Dolidze 39 & Dolidze 41, respectively.**